\documentclass[12pt]{iopart}

\usepackage{cite}

\begin{document}

\title{On the real matrix representation of PT-symmetric operators}
\author{Francisco M. Fern\'{a}ndez}

\address{INIFTA (UNLP, CCT La Plata-CONICET), Blvd. 113 y 64 S/N, Sucursal 4,
Casilla de Correo 16, 1900 La Plata,
Argentina}\ead{fernande@quimica.unlp.edu.ar}

\begin{abstract}
We discuss the construction of real matrix representations of PT-symmetric
operators. We show the limitation of a general recipe presented some time
ago for non-Hermitian Hamiltonians with antiunitary symmetry and propose a
way to overcome it. Our results agree with earlier ones for a particular
case.
\end{abstract}

\maketitle

\section{Introduction}

At first sight it is suprising that a subset of eigenvalues of a
complex-valued non-hermitian operator $\hat{H}$ can be real\cite{B07} (and
references therein). In order to provide a simple and general explanation of
this fact Bender et al\cite{BBM02} showed that it is possible to construct a
basis set of vectors so that the matrix representation of such operator is
real. As a result the secular determinant is real (the coefficients of the
characteristic polynomial are real) and its roots are either real or appear
in pairs of complex conjugate numbers. The argument is based on the
existence of an antiunitary symmetry $\hat{A}\hat{H}\hat{A}^{-1}=\hat{H}$,
where the antiunitary operator $\hat{A}$ satisfies $\hat{A}^{k}=\hat{1}$ for
$k$ odd. Bender et al\cite{BBM02} showed some illustrative examples of their
general result.

The procedure followed by Bender et al\cite{BBM02} for the construction of
the suitable basis set is reminiscent of the one used by Porter\cite{P65} in
the study of matrix representations of Hermitian operators. However, the
ansatz proposed by the latter author appears to be somewhat more general.

The purpose of this paper is to analyse the argument given by Bender et al%
\cite{BBM02} in more detail. In section~\ref{sec:Antiunitary} we outline the
main features of an antiunitary or antilinear operator and in section~\ref
{sec:Antiunitary_sym} we briefly discuss the concept of antiunitary
symmetry. In section~\ref{sec:real_matrix} we review the argument given by
Bender et al\cite{BBM02} and show that under certain conditions it does not
apply. We illustrate this point by means of the well known
harmonic-oscillator basis set and show how to overcome that shortcoming. In
section~\ref{sec:harmonic_osc} we discuss the harmonic-oscillator basis set
in more detail and in section~\ref{sec:conclusions} draw conclusions.

\section{Antiunitary operator}

\label{sec:Antiunitary}

As already mentioned above, a wide class of non-hermitian Hamiltonians with
unbroken PT symmetry exhibit real spectra\cite{B07}. In general, they are
invariant under an antilinear or antiunitary transformation of the form $%
\hat{A}^{-1}\hat{H}\hat{A}=\hat{H}$. The antiunitary operator $\hat{A}$
satisfies\cite{W60}
\begin{eqnarray}
\hat{A}\left( \left| f\right\rangle +\left| g\right\rangle \right) &=&\hat{A}%
\left| f\right\rangle +\hat{A}\left| g\right\rangle  \nonumber \\
\hat{A}c\left| f\right\rangle &=&c^{*}\hat{A}\left| f\right\rangle ,
\label{eq:antiunitary_1}
\end{eqnarray}
for any pair of vectors $\left| f\right\rangle $ and $\left| g\right\rangle $
and arbitrary complex number $c$, where the asterisk denotes complex
conjugation. This definition is equivalent to
\begin{equation}
\left\langle \hat{A}f\right. \left| \hat{A}g\right\rangle =\left\langle
f\right. \left| g\right\rangle ^{*}  \label{eq:antiuniary_2}
\end{equation}
One can easily derive the pair of equations (\ref{eq:antiunitary_1}) from (%
\ref{eq:antiuniary_2}) so that the latter can be considered to be the actual
definition of antiunitary operator\cite{W60}.

If $\hat{K}$ is an antilinear operator such that $\hat{K}^{2}=\hat{1}$ (for
example, the complex conjugation operator) then it follows from (\ref
{eq:antiuniary_2}) that $\hat{A}\hat{K}=\hat{U}$ is unitary ($\hat{U}%
^{\dagger }=\hat{U}^{-1}$); that is to say the inner product $\left\langle
f\right. \left| g\right\rangle $ remains invariant under $\hat{U}$:
\begin{equation}
\left\langle \hat{A}\hat{K}f\right. \left| \hat{A}\hat{K}g\right\rangle
=\left\langle \hat{K}f\right. \left| \hat{K}g\right\rangle ^{*}=\left\langle
f\right. \left| g\right\rangle
\end{equation}
In other words, any antilinear operator $\hat{A}$ can be written as a
product of a unitary operator and the complex conjugation operation\cite{W60}%
. Exactly in the same way we can easily prove that $\hat{A}^{2j}$ is unitary
and $\hat{A}^{2j+1}$ antiunitary.

In their discussion of real matrix representations of non-hermitian
Hamiltonians Bender et al\cite{BBM02} considered Hamiltonians $\hat{H}$ with
antiunitary symmetry
\begin{equation}
\hat{A}\hat{H}\hat{A}^{-1}=\hat{H}  \label{eq:Antiunitary_sym}
\end{equation}
where $\hat{A}$ satisfies the additional condition
\begin{equation}
\hat{A}^{2k}=\hat{1},\,k\;\mathrm{odd}  \label{eq:A^2k=1}
\end{equation}
Since $\hat{B}=\hat{A}^{k}$ is antiunitary and satisfies $\hat{B}^{2}=\hat{1}
$ we can restrict our discussion to the case $k=1$ without loss of
generality. Therefore, from now on we substitute the condition
\begin{equation}
\hat{A}^{2}=\hat{1}  \label{eq:A^2=1}
\end{equation}
for the apparently more general equation (\ref{eq:A^2k=1}). From now on we
refer to equation (\ref{eq:Antiunitary_sym}) as $A$-symmetry and to the
operator $\hat{H}$ as $A$-symmetric for short.

\section{Antiunitary symmetry}

\label{sec:Antiunitary_sym}

It follows from the antiunitary invariance (\ref{eq:Antiunitary_sym}) that $[%
\hat{H},\hat{A}]=0$. Therefore, if $\left| \psi \right\rangle $ is an
eigenvector of $\hat{H}$ with eigenvalue $E$
\begin{equation}
\hat{H}\left| \psi \right\rangle =E\left| \psi \right\rangle ,
\end{equation}
we have
\begin{equation}
\lbrack \hat{H},\hat{A}]\left| \psi \right\rangle =\hat{H}\hat{A}\left| \psi
\right\rangle -\hat{A}\hat{H}\left| \psi \right\rangle =\hat{H}\hat{A}\left|
\psi \right\rangle -E^{*}\hat{A}\left| \psi \right\rangle =0.
\end{equation}
This equation tells us that if $\left| \psi \right\rangle $ is eigenvector
of $\hat{H}$ with eigenvalue $E$ then $\hat{A}\left| \psi \right\rangle $ is
also eigenvector with eigenvalue $E^{*}$. That is to say: the eigenvalues
are either real or appear as pairs of complex conjugate numbers. In the
former case
\begin{equation}
\hat{H}\hat{A}\left| \psi \right\rangle =E\hat{A}\left| \psi \right\rangle ,
\label{eq:gen_unb_sym}
\end{equation}
that contains the condition of unbroken symmetry\cite{B07}
\begin{equation}
\hat{A}\left| \psi \right\rangle =\lambda \left| \psi \right\rangle
\label{eq:unb_sym}
\end{equation}
as a particular case. Note that equation (\ref{eq:gen_unb_sym}) applies to
the case in which $\hat{A}\left| \psi \right\rangle $ is a linear
combination of degenerate eigenvectors of $\hat{H}$ with eigenvalue $E$. An
illustrative example of this more general condition for real eigenvalues is
given elsewhere\cite{FG13}.

\section{Real matrix representation}

\label{sec:real_matrix}

Bender et al\cite{BBM02} put forward a straightforward procedure for
obtaining a basis set in which an $A$-symmetric Hamiltonian has a real
matrix representation. They proved that for an $A$-adapted basis set $%
\{\left| n_{A}\right\rangle \}$%
\begin{equation}
\hat{A}\left| n_{A}\right\rangle =\left| n_{A}\right\rangle
\label{eq:A-adapted}
\end{equation}
the matrix elements of the invariant Hamiltonian operator are real
\begin{equation}
\left\langle m_{A}\right| \hat{H}\left| n_{A}\right\rangle =\left\langle
m_{A}\right| \hat{H}\left| n_{A}\right\rangle ^{*}
\end{equation}
Those authors proposed to construct $\left| n_{A}\right\rangle $ as
(remember that we have restricted present discussion to $k=1$ without loss
of generality)
\begin{equation}
\left| n_{A}\right\rangle =\left| n\right\rangle +\hat{A}\left|
n\right\rangle  \label{eq:|nA>_Bender}
\end{equation}
where $\{\left| n\right\rangle \}$ is any orthonormal basis set.

It is not difficult to prove that this recipe does not apply to any basis
set. According to equation (\ref{eq:A^2=1}) we can find a basis set $%
\{\left| n,\sigma \right\rangle \}$ that satisfies
\begin{equation}
\hat{A}\left| n,\sigma \right\rangle =\sigma \left| n,\sigma \right\rangle
,\,\sigma =\pm 1  \label{eq:|n,s>_1}
\end{equation}
Consequently, all the vectors
\begin{equation}
\left| n,\sigma \right\rangle _{A}=\left| n,\sigma \right\rangle +\hat{A}%
\left| n,\sigma \right\rangle =\left( 1+\sigma \right) \left| n,\sigma
\right\rangle
\end{equation}
with $\sigma =-1$ vanish and the resulting $A$-adapted vector set is not
complete. We conclude that the basis set $\{\left| n\right\rangle \}$ should
be chosen carefully in order to apply the recipe of Bender et al\cite{BBM02}%
. In fact, the authors showed a particular example where it certainly
applies.

We can construct the basis set $\{\left| n,\sigma \right\rangle \}$ from any
orthonormal basis set $\{\left| n\right\rangle \}$ in the following way
\begin{equation}
\left| n,\sigma \right\rangle =N_{n,\sigma }\hat{Q}_{\sigma }\left|
n\right\rangle ,\,\hat{Q}_{\sigma }=\frac{1}{2}\left( 1+\sigma \hat{A}\right)
\label{eq:|n,s>_2}
\end{equation}
where $N_{n,\sigma }$ is a suitable normalization factor. It already
satisfies equation (\ref{eq:|n,s>_1}) because $\hat{A}\hat{Q}_{\sigma
}=\sigma \hat{Q}_{\sigma }$. In order to overcome the shortcoming in the
recipe (\ref{eq:|nA>_Bender}) we define the $A$-adapted basis set $%
B_{A}=\{\left| n_{A}^{+}\right\rangle =\left| n,1\right\rangle ,\left|
n_{A}^{-}\right\rangle =i\left| n,-1\right\rangle \}$. Note that the vectors
$\left| n_{A}^{\pm }\right\rangle $ satisfy the requirement (\ref
{eq:A-adapted}) and that $B_{A}$ is complete. In principle there is no
guarantee of orthogonality, but such a difficulty does not arise in the
examples discussed below.

The vectors $\left| v_{n}^{\pm }\right\rangle =\frac{1}{\sqrt{2}}\left(
\left| n_{A}^{+}\right\rangle \pm \left| n_{A}^{-}\right\rangle \right) $
also satisfy the requirement (\ref{eq:A-adapted}) and in the particular case
of a two-dimensional space they lead to the $A$-adapted basis set chosen by
Bender et al\cite{BBM02} to introduce the issue by means of a simple example
.

As a particular case consider the parity-time antiunitary operator $\hat{A}=%
\hat{P}\hat{T}$, where $\hat{P}$ an $\hat{T}$ are the parity and
time-reversal operators, respectively\cite{P65}. Let $\{\left|
n\right\rangle ,\,n=0,1,\ldots \}$ be the basis set of eigenvectors of the
Harmonic oscillator $\hat{H}_{0}=\hat{p}^{2}+\hat{x}^{2}$ that are real and
satisfy $\hat{P}\left| n\right\rangle =(-1)^{n}\left| n\right\rangle $ so
that $\hat{A}\left| n\right\rangle =(-1)^{n}\left| n\right\rangle $. It is
clear that the recipe (\ref{eq:|nA>_Bender}) does not apply to this simple
case. On the other hand, present recipe yields the $A$-adapted basis set $%
B_{A}^{HO}=\{\left| 2n\right\rangle ,i\left| 2n+1\right\rangle
,\,n=0,1,\ldots \}$ which is obviously complete.

Every vector of the orthonormal basis set $B_{A}^{HO}$ in the coordinate
representation can be expressed as a linear combination of the elements of
the nonorthogonal basis set $\{f_{n}(x)=e^{-x^{2}/2}(ix)^{n},\,n=0,1,\ldots
\}$. By means of a slight generalization of the latter Znojil\cite{Z99}
derived a recurrence relation with real coefficients for a family of complex
anharmonic potentials. This author also constructed a real matrix
representation of a PT-symmetric oscillator in terms of the eigenvectors of $%
\hat{A}=\hat{P}\hat{T}$\cite{Z02}. Note that his vectors $\left|
S_{n}\right\rangle $ and $\left| L_{n}\right\rangle $ are our $\left|
n,1\right\rangle $ and $\left| n,-1\right\rangle $ respectively.

Following Porter\cite{P65} we can try the ansatz
\begin{equation}
\left| n_{A}\right\rangle =a_{n}\left| n\right\rangle +\hat{A}a_{n}\left|
n\right\rangle =a_{n}\left| n\right\rangle +a_{n}^{*}\hat{A}\left|
n\right\rangle  \label{eq:|nA>_Porter}
\end{equation}
that already satisfies $\hat{A}\left| n_{A}\right\rangle =\left|
n_{A}\right\rangle $. This definition of $A$-adapted basis set is slightly
more general than equation (\ref{eq:|nA>_Bender}). When $\hat{A}\left|
n\right\rangle =(-1)^{n}\left| n\right\rangle $ we simply choose $a_{n}=%
\frac{1}{2}(1+i)$ and obtain the result above for the particular case of the
harmonic-oscillator basis set. Note that the resulting expressions (we can
also choose $a_{n}=\frac{1}{2}(1-i)$) are similar to those in equation (16)
in the paper of Bender et al\cite{BBM02}.

\section{The harmonic-oscillator basis set}

\label{sec:harmonic_osc}

Many examples of $PT$-symmetric Hamiltonians are one-dimensional models of
the form\cite{B07}
\begin{equation}
\hat{H}=\hat{p}^{2}+V(x),
\end{equation}
where
\begin{equation}
V(-x)^{*}=V(x).
\end{equation}
We can write $V(x)$ as the sum of its even $V_{e}(-x)=V_{e}(x)$ and odd $%
V_{o}(-x)=-V_{o}(x)$ parts
\begin{equation}
V(x)=V_{e}(x)+V_{o}(x),
\end{equation}
where
\begin{eqnarray}
V_{e}(x) &=&\frac{1}{2}\left[ V(x)+V(-x)\right] =\Re V(x),  \nonumber \\
V_{o}(x) &=&\frac{1}{2}\left[ V(x)-V(-x)\right] =i\Im V(x).
\end{eqnarray}

For convenience we change the notation of the preceding section and define
the $A$-adapted basis set $\{\varphi _{n}\}$ as
\begin{eqnarray}
\left| \varphi _{2n}\right\rangle  &=&\left| 2n\right\rangle   \nonumber \\
\left| \varphi _{2n+1}\right\rangle  &=&i\left| 2n+1\right\rangle
,\,n=0,1,\ldots ,  \label{eq:phi->varphi}
\end{eqnarray}
where $\left\{ \left| n\right\rangle \right\} $ is the harmonic-oscillator
basis set. Therefore
\begin{eqnarray}
\left\langle \varphi _{2n}\right| \hat{p}^{2}\left| \varphi
_{2m}\right\rangle  &=&\left\langle 2n\right| \hat{p}^{2}\left|
2m\right\rangle   \nonumber \\
\left\langle \varphi _{2n}\right| \hat{p}^{2}\left| \varphi
_{2m+1}\right\rangle  &=&\left\langle 2n\right| \hat{p}^{2}\left|
2m+1\right\rangle =0  \nonumber \\
\left\langle \varphi _{2m+1}\right| \hat{p}^{2}\left| \varphi
_{2n}\right\rangle  &=&\left\langle 2m+1\right| \hat{p}^{2}\left|
2n\right\rangle =0  \nonumber \\
\left\langle \varphi _{2n+1}\right| \hat{p}^{2}\left| \varphi
_{2m+1}\right\rangle  &=&\left\langle 2n+1\right| \hat{p}^{2}\left|
2m+1\right\rangle
\end{eqnarray}

and
\begin{eqnarray}
\left\langle \varphi _{2n}\right| V\left| \varphi _{2m}\right\rangle
&=&\left\langle 2n\right| \Re V\left| 2m\right\rangle   \nonumber \\
\left\langle \varphi _{2n+1}\right| V\left| \varphi _{2m}\right\rangle
&=&\left\langle 2n+1\right| \Im V\left| 2m\right\rangle   \nonumber \\
\left\langle \varphi _{2n}\right| V\left| \varphi _{2m+1}\right\rangle
&=&-\left\langle 2n\right| \Im V\left| 2m+1\right\rangle   \nonumber \\
\left\langle \varphi _{2n+1}\right| V\left| \varphi _{2m+1}\right\rangle
&=&\left\langle 2n+1\right| \Re V\left| 2m+1\right\rangle
\end{eqnarray}
It is clear that all the matrix elements $H_{mn}=\left\langle \varphi
_{m}\right| \hat{H}\left| \varphi _{n}\right\rangle $ are real and the basis
is complete since
\begin{equation}
\sum_{n}\left| \varphi _{n}\right\rangle \left\langle \varphi _{n}\right|
=\sum_{n}\left| n\right\rangle \left\langle n\right| =\hat{1}
\end{equation}
Besides, the matrix representation of the Hamiltonian operator in the basis
set discussed above
\begin{equation}
\hat{H}=\sum_{m}\sum_{n}\left| \varphi _{m}\right\rangle \left\langle
\varphi _{m}\right| \hat{H}\left| \varphi _{n}\right\rangle \left\langle
\varphi _{n}\right|
\end{equation}
is similar to the one proposed by Znojil\cite{Z02} some time ago.

The unitary basis transformation (\ref{eq:phi->varphi}) is given by the
unitary operator
\begin{equation}
\hat{U}=\sum_{n=0}^{\infty }\left( \left| 2n\right\rangle \left\langle
2n\right| +i\left| 2n+1\right\rangle \left\langle 2n+1\right| \right)
\end{equation}
that satisfies $\hat{U}^{\dagger }=\hat{U}^{*}=\hat{T}\hat{U}\hat{T}$ and $%
\hat{U}^{2}=\hat{P}$. If $\mathbf{H}$ and $\mathbf{U}$ are the martix
representations of the operators $\hat{H}$ and $\hat{U}$, respectively, in
the basis set $\{\left| n\right\rangle \}$ and $\mathbf{I}$ is the identity
matrix, then the secular determinant $\left| \mathbf{H}-E\mathbf{I}\right| =$
$\left| \mathbf{U}\left( \mathbf{H}-E\mathbf{I}\right) \mathbf{U}^{\dagger
}\right| =$ $\left| \mathbf{UHU}^{\dagger }-E\mathbf{I}\right| $ is real
because the matrix elements of $\mathbf{UHU}^{\dagger }$ are all real. This
result applies even to the approximate finite matrix representations of
operators appearing in the diagonalization method\cite{BW12,FG13}. As a
consequence, the coefficients of the characteristic polynomial are real and
their roots are either real or complex conjugate numbers.

\section{Conclusions}

\label{sec:conclusions}

We have shown that the recipe proposed by Bender et al\cite{BBM02} for the
construction of real matrix representations of $A$-symmetric Hamiltonians
may fail under certain conditions, for example, when $\hat{A}\left|
n\right\rangle =(-1)^{n}$ $\left| n\right\rangle $. In this case one can
easily construct an $A$-adapted basis set as $\left| n_{A}\right\rangle
=i^{n}\left| n\right\rangle $ that is complete and satisfies the required
condition $\hat{A}\left| n_{A}\right\rangle =\left| n_{A}\right\rangle $.
One of the most commonly used basis set, the harmonic-oscillator one,
already belongs to this class. There is no unique way of constructing the $A$%
-adapted basis set; for example, the ansatz proposed by Porter\cite{P65} (in
the form outlined above in section \ref{sec:real_matrix}) yields basically
the same basis vectors except for the phase factors.

\end{document}